\documentclass[12pt,reqno]{article}
\usepackage{amsfonts}
\usepackage{amsmath}
\usepackage{amsbsy}

\usepackage{amssymb,latexsym}
\usepackage{setspace}
 \numberwithin{equation}{section}
\numberwithin{equation}{section}

\begin{document}
 \allowdisplaybreaks[1]
\title{An Implicit Decoupling for the Dilatons and the Axions of the Heterotic String}
\author{Nejat T. Y$\i$lmaz\\
          Department of Mathematics
and Computer Science,\\
\c{C}ankaya University,\\
\"{O}\u{g}retmenler Cad. No:14,\quad  06530,\\
 Balgat, Ankara, Turkey.\\
          \texttt{ntyilmaz@cankaya.edu.tr}}
\maketitle
\begin{abstract}A set of consistency conditions is derived for
the massless sector of the $D$-dimensional $E_{8}\times E_{8}$
heterotic string. Under the solvable Lie algebra gauge these
conditions are further formulated explicitly in terms of the
dilatons and the axions. It is then shown that these consistency
conditions which are satisfied by the solution space give way to
an implicit decoupling between the coset scalar sector namely the
dilatons and the axions and the gauge fields of the theory.
\end{abstract}

\section{Introduction}
The global (rigid) symmetry of the scalar sectors of the
supergravity theories can be extended to be the global symmetry of
the entire bosonic sector of the theory. The global symmetries of
the supergravities give us the non-perturbative U-duality
symmetries of the relative string theories \cite{kiritsis,
nej125,nej126}. Therefore understanding the role of the scalars in
the supergravity dynamics has implications not only on its own
right but it will also help us to understand the duality nature of
the string theories.

The ten dimensional $\mathcal{N=}1$ type I supergravity theory
which is coupled to the Yang-Mills theory \cite{d=10,tani15} is
the low energy effective limit or the massless sector of the type
I superstring theory and the heterotic string theory
\cite{kiritsis,heterotic}. When the coupling Yang-Mills multiplet
number is $N=16$, the $D=10$ Yang-Mills supergravity has the
$E_{8}\times E_{8}$ gauge symmetry thus it corresponds to the low
energy effective limit of the $E_{8}\times E_{8}$ ten dimensional
heterotic string theory which forms the massless background
coupling \cite{kiritsis}. Due to the general Higgs vacuum
structure the full symmetry $E_{8}\times E_{8}$ can be broken down
to its maximal torus subgroup $U(1)^{16}$. In this case we have
the low energy effective theory of the fully Higgsed ten
dimensional $E_{8}\times E_{8}$ heterotic string. In
\cite{heterotic} the Kaluza-Klein compactification of the bosonic
sector of the ten dimensional $\mathcal{N=}1$ simple supergravity
that is coupled to $N$ abelian gauge multiplets on the tori
$T^{10-D}$ is performed and the $D$-dimensional fully Higgsed
effective bosonic heterotic string lagrangian is obtained. It is
also shown in the same work that when a single scalar is decoupled
from the others, the rest of the scalars of the $D$-dimensional
theory can be formulated as $G/K$ symmetric space sigma model.
Since the global symmetry groups of the compactified
$D$-dimensional effective heterotic string are non-compact real
forms one can make use of the solvable Lie algebra gauge
\cite{fre} to parametrize the scalar cosets and to construct the
scalar lagrangians. In such a gauge simply the gauge forming
solvable Lie algebra is composed of certain Cartan and positive
root generators of the Lie algebra of $G$ \cite{hel}.

In this work starting from the $D$-dimensional effective bosonic
heterotic string lagrangian derived in \cite{heterotic} we show
that the field equation of the single decoupled dilaton generates
a set of consistency conditions which are satisfied by the
elements of the solution space of the theory. We interpret these
hidden consistency conditions as an implicit decoupling between
the $G/K$ coset scalars and the $U(1)$ gauge fields. Therefore we
also prove that the coset scalar solution space of the
$D$-dimensional effective heterotic string is embedded in the
solution space of the pure $G/K$ non-linear sigma model.
\section{Consistency Conditions and the Scalar Matter Decoupling}\label{section23}
In this section starting from the bosonic lagrangian of the
$D$-dimensional fully Higgsed low energy effective heterotic
string \cite{heterotic} we will show that the field equation of
the decoupled dilaton leads to two sets of consistency conditions
which are satisfied by the solution space of the theory. Although
these conditions are in general valid for generic coset scalar
fields which are defined without specifying a gauge for the coset
parametrization we will choose the solvable Lie algebra
parametrization \cite{fre} to further simplify these conditions.
By assuming a matrix representation for the Lie algebra of the
global symmetry group we will express the general consistency
conditions in terms of the dilatons and the axions of the theory.
Then we will show that these consistency conditions reveal a
hidden decoupling of the gauge fields in the field equations of
the coset scalars although there exists a scalar-matter coupling
term in the lagrangian.

The bosonic lagrangian of the $D=10$, $\mathcal{N}=1$ abelian
Yang-Mills supergravity which is coupled to $N$ $U(1)$ gauge
multiplets can be given as \cite{d=10,tani15,heterotic}
\begin{equation}\label{cs21}
{\mathcal{L}}_{10}=R\ast1-\frac{1}{2}\ast d\phi_{1}\wedge
d\phi_{1}-\frac{1}{2}e^{\phi_{1}}\ast F_{(3)}\wedge F_{(3)}
-\frac{1}{2}e^{\frac{1}{2}\phi_{1}}\sum\limits_{I=1}^{N}\ast
G_{(2)}^{I}\wedge G_{(2)}^{I},
\end{equation}
where $G_{(2)}^{I}=dB_{(1)}^{I}$ are the $N$ $U(1)$ gauge field
strengths. In \eqref{cs21} $\phi_{1}$ is a scalar field and we
have a two-form field $A_{(2)}$ whose field strength is defined as
\begin{equation}\label{cs22}
F_{(3)}=dA_{(2)}+\frac{1}{2}B_{(1)}^{I}\wedge d B_{(1)}^{I}.
\end{equation}
When the bosonic sector of the ten dimensional simple
$\mathcal{N=}1$ supergravity which is coupled to $N$ abelian gauge
multiplets is compactified on the Euclidean torus $T^{10-D}$ in
$D$-dimensions one obtains the reduced lagrangian \cite{heterotic}
\begin{subequations}\label{cs23}
\begin{align}
{\mathcal{L}}_{D}&=R\ast1-\frac{1}{2}\ast d\phi\wedge
d\phi+\frac{1}{4}tr(\ast d{\mathcal{M}}^{-1}\wedge
d{\mathcal{M}})\notag\\
\notag\\
&\quad -\frac{1}{2}e^{-\sqrt{8/(D-2)}\phi}\ast F_{(3)}\wedge
F_{(3)}-\frac{1}{2}e^{-\sqrt{2/(D-2)}\phi}\ast
H_{(2)}^{T}\wedge{\mathcal{M}}H_{(2)},\tag{\ref{cs23}}
\end{align}
\end{subequations}
with
\begin{equation}\label{cs24}
H_{(2)}=dC_{(1)},
\end{equation}
where $C_{(1)}$ is a column vector of one-form fields whose
dimension is $(20-2D+N)$. The field strength $F_{(3)}$ in
\eqref{cs23} is defined as
\begin{equation}\label{cs25}
F_{(3)}=dA_{(2)}+\frac{1}{2}C_{(1)}^{T}\:\Omega\: dC_{(1)}.
\end{equation}
The definitions of the potentials used in writing \eqref{cs23} in
terms of the original Kaluza-Klein potentials which emerge from
the reduction ansatz used in \eqref{cs21} can be found in
\cite{heterotic}. Apart form the single dilaton $\phi$ which is
decoupled from the rest the scalars in \eqref{cs23} parametrize
the coset
\begin{equation}\label{cs25.5}
\frac{O(10-D+N,10-D)}{O(10-D+N)\times O(10-D)}.
\end{equation}
The $(20-2D+N)\times(20-2D+N)$ matrix $\Omega$ in \eqref{cs25} is
\begin{equation}\label{cs26}
\Omega=\left(\begin{array}{ccc}
  0 & 0 & -{\mathbf{1}}_{(10-D)} \\
  0 & {\mathbf{1}}_{(N)} & 0 \\
  -{\mathbf{1}}_{(10-D)} & 0 & 0 \\
\end{array}\right),
\end{equation}
where ${\mathbf{1}}_{(n)}$ is the $n\times n$ unit matrix.
$\Omega$ is the invariant metric of $O(10-D+N,10-D)$. The scalar
sector lagrangian in \eqref{cs23} is based on the internal metric
\begin{equation}\label{cs27}
\mathcal{M}=\nu ^{T}\nu ,
\end{equation}
where $\nu$ is the $O(10-D+N,10-D)/O(10-D+N)\times O(10-D)$ coset
representative. One can use the solvable Lie algebra
parametrization \cite{fre,hel}
\begin{equation}\label{cs28}
\nu=e^{\frac{1}{2}\phi ^{i}H_{i}}e^{\chi ^{m}E_{m}},
\end{equation}
to parametrize the coset representative $\nu$. This
parametrization is a result of the Iwasawa decomposition
\begin{equation}\label{cs29}
o(10-D+N,10-D)=k_{0}\oplus s_{0},
\end{equation}
where $k_{0}$ is the Lie algebra of $O(10-D+N)\times O(10-D)$
which is a maximal compact subgroup of $O(10-D+N,10-D)$ and
$s_{0}$ is a solvable Lie subalgebra of $o(10-D+N,10-D)$
\cite{hel,nej2}. The set $\{H_{i}\}$ which is composed of a
certain number of Cartan generators and the set $\{E_{m}\}$ which
is composed of a certain number of positive root generators are
the generators of $s_{0}$ \cite{nej2}. The $(10-D)\times(10-D+N)$
coset scalars are divided into the dilatons $\{\phi^{i}\}$ and the
axions $\{\chi^{m}\}$.

As we have mentioned in the introduction in this work our aim is
to derive a set of consistency conditions which reveal an implicit
decoupling in the scalar-matter structure of \eqref{cs23}. We
should state that for observing the above mentioned consistency
conditions we do not have to inspect the entire set of field
equations of \eqref{cs23}. It will be sufficient to take a look at
only the field equation of the decoupled dilaton $\phi$ to obtain
the necessary consistency conditions. Now if we vary the
lagrangian \eqref{cs23} with respect to the scalar field $\phi$ we
obtain the corresponding field equation as
\begin{equation}\label{cs210}
\begin{aligned}
(-1)^{D}d(\ast
d\phi)&=\frac{1}{2}\sqrt{8/(D-2)}\:e^{-\sqrt{\frac{8}{(D-2)}}\phi}
\ast F_{(3)}\wedge
F_{(3)}\\
\\
&\quad
+\frac{1}{2}\sqrt{2/(D-2)}\:e^{-\sqrt{\frac{2}{(D-2)}}\phi}\mathcal{M}_{ij}
\ast H_{(2)}^{i}\wedge H_{(2)}^{j}.
\end{aligned}
\end{equation}
This equation can be written as
\begin{equation}\label{cs211}
-\sqrt{2/(D-2)}\mathcal{L}_{mat}=(-1)^{D}d(\ast
d\phi)-\frac{1}{2}\sqrt{8/(D-2)}\:e^{-\sqrt{\frac{8}{(D-2)}}\phi}
\ast F_{(3)}\wedge F_{(3)},
\end{equation}
where
\begin{equation}\label{cs212}
\mathcal{L}_{mat}=-\frac{1}{2}\:e^{-\sqrt{\frac{2}{(D-2)}}\phi}\mathcal{M}_{ij}
\ast H_{(2)}^{i}\wedge H_{(2)}^{j},
\end{equation}
is the term which couples the coset scalars to the matter gauge
fields in \eqref{cs23}. Now we should observe that since the right
hand side of \eqref{cs211} does not depend on the dilatons and the
axions if we take the partial derivative of both sides of
\eqref{cs211} with respect to the dilatons $\{\phi^{i}\}$ and the
axions $\{\chi^{m}\}$ we obtain
\begin{equation}\label{cs214}
\frac{\partial\mathcal{L}_{mat}}{\partial\phi^{i}}=\frac{\partial\mathcal{L}_{mat}}{\partial\chi^{m}}=0.
\end{equation}
Since in deriving \eqref{cs214} we have started from the field
equation \eqref{cs210} these conditions are consistency conditions
which are satisfied by the solution space of \eqref{cs23}. Before
discussing their implications in the scalar sector we will work on
them further more and derive the explicit form of the conditions
\eqref{cs214}. For any matrix function $\omega(x)$ we have
\cite{sat,hall}
\begin{equation}\label{cs215}
\frac{\partial e^{\omega}}{\partial
x}=e^{\omega}(\omega^{\prime}-\frac{1}{2!}[\omega,\omega^{\prime}]+\frac{1}{3!}[\omega,[\omega,
\omega^{\prime}]]-\cdot\cdot\cdot),
\end{equation}
with
\begin{equation}\label{cs216}
 \omega^{\prime}\equiv\frac{\partial\omega}{\partial x}.
\end{equation}
When we choose a ($20-2D+N$)-dimensional representation for
$o(10-D+N,10-D)$ and express the generators $\{H_{i}\}$ and
$\{E_{m}\}$ as matrices bearing in mind that the generators
$\{H_{i}\}$ are Cartan generators \cite{nej2} after some algebra
one can show that
\begin{equation}\label{cs217}
\frac{\partial\nu}{\partial\phi^{i}}=\frac{1}{2}H_{i}\nu\quad\text{and}\quad
\frac{\partial\nu^{T}}{\partial\phi^{i}}=\frac{1}{2}\nu^{T}H_{i}^{T}.
\end{equation}
Therefore we find
\begin{equation}\label{cs218}
\frac{\partial\mathcal{M}}{\partial\phi^{i}}=\frac{1}{2}(\mathcal{A}+\mathcal{A}^{T}),
\end{equation}
where we define
\begin{equation}\label{cs219}
\mathcal{A}=\nu^{T}H_{i}\nu.
\end{equation}
Now from \eqref{cs215} we also define
\begin{equation}\label{cs220}
\begin{aligned}
O(E_{m})\equiv\frac{\partial
(e^{\chi^{n}E_{n}})}{\partial\chi^{m}}&=e^{\chi^{n}E_{n}}(E_{m}-\frac{1}{2!}[\chi^{n}E_{n},E_{m}]\\
\\
&\quad+\frac{1}{3!}[\chi^{n}E_{n} ,[\chi^{t}E_{t},
E_{m}]]-\cdot\cdot\cdot),
\end{aligned}
\end{equation}
where we have used
\begin{equation}\label{cs221}
\frac{\partial (\chi^{n}E_{n})}{\partial\chi^{m}}=E_{m}.
\end{equation}
After some algebra one can prove that
\begin{equation}\label{cs222}
 O^{T}(E_{m})=O(E_{m}^{T}).
\end{equation}
By using the definition \eqref{cs220} and also the identity
\eqref{cs222} similarly for the axions we find that
\begin{equation}\label{cs223}
\frac{\partial\mathcal{M}}{\partial\chi^{m}}=\mathcal{B}+\mathcal{B}^{T},
\end{equation}
where we introduce
\begin{equation}\label{cs224}
\mathcal{B}=\nu^{T}e^{\frac{1}{2}\phi ^{i}H_{i}}O(E_{m}).
\end{equation}
We should state that since $\{E_{m}\}$ are the generators of a
nilpotent Lie subalgebra of $o(10-D+N,10-D)$ \cite{nej2} one can
prove that the series in \eqref{cs220} should terminate after a
finite number of terms \cite{carter,onis}. Thus the calculation of
\eqref{cs224} is a straightforward task after choosing the
representation. Now from \eqref{cs214} we have
\begin{equation}\label{cs225}
\begin{aligned}
 \frac{\partial\mathcal{L}_{mat}}{\partial\phi^{i}}&=-
 \frac{1}{2}\:e^{-\sqrt{\frac{2}{(D-2)}}\phi}\frac{\partial\mathcal{M}_{kl}}{\partial\phi^{i}}
\ast H_{(2)}^{k}\wedge H_{(2)}^{l}=0,\\
\\
\frac{\partial\mathcal{L}_{mat}}{\partial\chi^{m}}&=-
 \frac{1}{2}\:e^{-\sqrt{\frac{2}{(D-2)}}\phi}\frac{\partial\mathcal{M}_{ij}}{\partial\chi^{m}}
\ast H_{(2)}^{i}\wedge H_{(2)}^{j}=0.
\end{aligned}
\end{equation}
By using \eqref{cs218} and \eqref{cs223} also by further
simplifying we can finally write the consistency conditions in
\eqref{cs225} as
\begin{equation}\label{cs226}
\begin{aligned}
 \mathcal{A}_{ij}\ast H_{(2)}^{i}\wedge H_{(2)}^{j}&=0,\\
\\
\mathcal{B}_{ij}\ast H_{(2)}^{i}\wedge H_{(2)}^{j}&=0.
\end{aligned}
\end{equation}
Before concluding we should discuss how these conditions bring out
an implicit decoupling between the coset scalars namely the
dilatons and the axions and the $U(1)$ gauge fields. The scalar
sector which governs the $(10-D)\times(10-D+N)$ coset scalars
$\{\phi^{i}\}$ and $\{\chi^{m}\}$ in \eqref{cs23} can be given as
\begin{equation}\label{cs227}
\mathcal{L}(\phi^{i},\chi^{m})=\frac{1}{4}tr( \ast
d\mathcal{M}^{-1}\wedge d\mathcal{M})+\mathcal{L}_{mat}.
\end{equation}
Now if we vary this lagrangian with respect to the dilatons
$\{\phi^{i}\}$ and the axions $\{\chi^{m}\}$ to find the
corresponding field equations we immediately see that since the
matter lagrangian in \eqref{cs227} does not depend on the field
strengths of the dilatons and the axions and moreover due to the
conditions \eqref{cs214} which are satisfied by the elements of
the solution space the dilaton and the axion field equations are
equivalent to the ones which would be obtained by directly varying
the pure $G/K$ coset sigma model lagrangian
\begin{equation}\label{cs228}
\mathcal{L}_{pure scalar}=\frac{1}{4}tr( \ast
d\mathcal{M}^{-1}\wedge d\mathcal{M}).
\end{equation}
Thus we prove that the coset scalar solutions of the field
equations of \eqref{cs23} are contained in the general solution
space of the pure symmetric space sigma model lagrangian
\eqref{cs228}\footnote{To see how the field equations of the pure
symmetric space sigma model can be derived one may refer to
\cite{nej2,ker2}.}. This may be identified as an implicit
decoupling structure of the coset scalars and the abelian gauge
fields $C_{(1)}$. Therefore we conclude that although there is a
coupling between the coset scalars and the abelian gauge fields at
the lagrangian level in \eqref{cs23} the consistency conditions we
have derived in \eqref{cs214} reflect an implicit decoupling
between the coset scalars and the abelian gauge fields in the
coset scalar field equations. However we should state that the
coset scalars and the rest of the fields are still coupled to each
other in the other field equations. Thus as a result we observe
that the dilaton and the axion solutions of the field equations of
the $D$-dimensional lagrangian \eqref{cs23} form up a subset of
the pure scalar coset sigma model solution space.
\section{Conclusion}
By inspecting the decoupled dilaton field equation of the bosonic
lagrangian of the $D$-dimensional fully Higgsed low energy
effective heterotic string \cite{heterotic} we have derived two
sets of consistency conditions which are satisfied by the elements
of the bosonic solution space. Assuming the solvable Lie algebra
parametrization for the scalar coset manifold and a matrix
representation for the global symmetry algebra we have obtained
the explicit forms of these conditions in terms of the dilatons
and the axions. Furthermore we have discussed that these
conditions bring out a hidden decoupling between the coset scalars
and the gauge fields of the theory from the scalar solution space
point of view despite the existence of a coupling term in the
lagrangian. Therefore we have revealed an implicit constraint
which shapes the bosonic solution space of the fully Higgsed
effective heterotic string.

The consistency conditions we have obtained can be effectively
used to generate ansatz for solving the bosonic field equations of
the theory. For this reason we have also derived the explicit form
of these conditions in the solvable Lie algebra gauge of the
scalar coset manifold. Besides we have shown that the coset scalar
solutions of the $D$-dimensional fully Higgsed low energy
effective heterotic string are embedded in the solution space of
the pure symmetric space $G/K$ sigma model. Thus the general
solutions of the pure coset scalar sector can also be used as
ansatz in solving the bosonic field equations of the
$D$-dimensional effective heterotic string.

The main achievement of this work is to reveal the fact that the
single dilaton which behaves distinctively in the $D$-dimensional
effective heterotic string lagrangian generates a scalar-matter
decoupling structure for the content of the scalar solution space
and its existence contributes an important set of consistency
conditions on the entire bosonic solution space. Such an implicit
decoupling of the coset scalars from the rest of the bosonic field
content also carries implications why the global symmetry of the
pure coset scalar sector can be extended to be the global symmetry
of the entire bosonic sector. Therefore this work may help us to
understand the global symmetries of the supergravities thus the
duality scheme of the string theory better.

Although we have assumed the solvable Lie algebra gauge to
parametrize the scalar coset manifold of the $D$-dimensional
theory the consistency conditions we have derived are valid for
any parametrization, scalar field definition and the formulation
of the coset scalar sector lagrangian. Therefore these conditions
can be studied further more in different formulations of the
scalar sector. We should state that the decoupling studied in this
work is a pseudo one instead since when one considers the rest of
the bosonic field equations of the theory one sees that the coset
scalars take part in the equations. Thus the solution
configuration of the coset scalars do depend on the other fields
since they couple to the rest of the field content in the other
field equations. However the field equations of the coset scalars
can be decoupled from the rest of the fields denoting that the
coset scalar solution space is contained in the pure scalar sector
non-linear sigma model solution space. With the perspective of
understanding the global symmetry structure of the supergravities
we have focussed on the role of the scalar coset sigma model in
the bosonic theory. For this reason we have inspected its bosonic
couplings and we have omitted the fermionic sector. One may do a
similar search to derive consistency conditions which would reveal
the relations between the pure scalar sector and its coupling
extensions in the entire theory.

\end{document}